\documentclass{article}

\usepackage[utf8]{inputenc}
\usepackage{graphicx}
\usepackage{amsmath}
\usepackage{listings}
\usepackage{subcaption}
\usepackage{amsfonts}
\usepackage{comment}
\usepackage{fullpage}
\usepackage{authblk}

\usepackage{hyperref}
\hypersetup{colorlinks, citecolor=blue}

\usepackage[
    backend=biber,
    style=numeric,
    sortcites=true,
    sorting=none,
    backref,
]{biblatex}

\title{Hierarchical Jacobi Iteration for Structured Matrices on GPUs using Shared Memory}

\author[,a]{Mohammad Shafaet Islam\thanks{Corresponding author \vfill \ \ \textit{Email addresses}: \texttt{moislam@mit.edu} (Mohammad Shafaet Islam), \texttt{qiqi@mit.edu} (Qiqi Wang)}}
\author[a]{Qiqi Wang}

\affil[a]{\footnotesize{\textit{Department of Aeronautics and Astronautics, Massachusetts Institute of Technology, 77 Massachusetts Avenue, Cambridge, MA 02139, USA}}}

\date{}

\begin{document}

\maketitle

\begin{abstract}
High fidelity scientific simulations modeling physical phenomena typically require solving large linear systems of equations which result from discretization of a partial differential equation (PDE) by some numerical method. This step often takes a vast amount of computational time to complete, and therefore presents a bottleneck in simulation work. Solving these linear systems efficiently requires the use of massively parallel hardware with high computational throughput, as well as the development of algorithms which respect the memory hierarchy of these hardware architectures to achieve high memory bandwidth. 

In this paper, we present an algorithm to accelerate Jacobi iteration for solving structured problems on graphics processing units (GPUs) using a hierarchical approach in which multiple iterations are performed within on-chip shared memory every cycle. A domain decomposition style procedure is adopted in which the problem domain is partitioned into subdomains whose data is copied to the shared memory of each GPU block. Jacobi iterations are performed internally within each block's shared memory, avoiding the need to perform expensive global memory accesses every step. We test our algorithm on the linear systems arising from discretization of Poisson’s equation in 1D and 2D, and observe speedup in convergence using our shared memory approach compared to a traditional Jacobi implementation which only uses global memory on the GPU. We observe a x8 speedup in convergence in the 1D problem and a nearly x6 speedup in the 2D case from the use of shared memory compared to a conventional GPU approach.
\end{abstract}

\section{Introduction}

Graphics processing units (GPUs) have emerged as a popular hardware architecture for scientific computing. While originally introduced for real-time graphics rendering in the gaming industry, GPUs were found to offer the computational horsepower and high memory bandwidth necessary to accelerate large scale computations which have traditionally been performed using CPUs. GPU computing has permeated many fields such as finance \cite{Finance2010}, computational chemistry \cite{Chemistry2013} and fluid dynamics \cite{GPUGems2004}.  

Simulation work often requires approximating the solution to a partial differential equation (PDE) governing some physical phenomenon in an efficient manner. This usually involves applying some numerical method to reduce the PDE into a set of linear equations. Solving the resulting linear system is often the bottleneck of many simulations. Improving the speed of linear solvers is essential to improving the value of simulation work. For this reason, there has been a tremendous effort to develop efficient GPU implementations of linear solver algorithms. As some examples, conjugate gradient solvers for sparse unstructured matrices and a multigrid solver for regular grids using GPU textures were developed by Bolz et al. \cite{Bolz2003}. To improve the performance of preconditioned conjugate gradient and GMRES, Li and Saad developed an efficient sparse matrix vector product on the GPU which exhibited up to eight times speedup relative to a parallel CPU implementation on several matrices from the SuiteSparse matrix collection \cite{Li2013}.

Jacobi iteration has been a popular iterative method of study on the GPU. While slow to converge compared to Krylov methods, Jacobi iteration is completely parallel making it quite amenable to GPU implementation \cite{Golub2007}. Furthermore, Jacobi iteration and other similar stationary iterative methods form the backbone of highly effective geometric and algebraic multigrid methods \cite{Briggs2000}. A number of studies have been conducted to assess the performance of Jacobi iteration on the GPU. Zhang et al. study the performance of a CPU and CUDA based GPU implementation of Jacobi iteration on a dense matrix. They observe a speedup of 19 times in double precision (59 times in floating point precision) for a matrix size of 7680 (when all GPU threads are operating) compared to the CPU performance \cite{Zhang2009}. Wang et al. perform studies of Jacobi iteration on sparse matrices on the GPU. Their implementation uses shared memory to store the solution at each step, and they observe a speedup factor of up to 100 times relative to their CPU implementation \cite{Wang2009}. Amorim et al. apply a weighted version of Jacobi iteration to a structured matrix resulting from a finite element discretization of a set of PDEs from cardiac modeling. They compare the performance of an OpenGL implementation to several CUDA implementations with global coalesced memory accesses, shared memory and texture memory, and find that a CUDA implementation with texture memory yields the best performance \cite{Amorim2009}. Jacobi iteration on a 2D Laplace equation is studied by Cecilia et al. They measure the time to perform a set number of iterations in CUDA using several code optimizations which involve the use of shared memory, and observe a three to four times speedup for various domain sizes compared to their unoptimized CUDA implementation (and even higher speedups relative to their CPU implementation) \cite{Cecilia2012}. Several multi-CPU, multi-GPU and hybrid CPU/GPU implementations of Jacobi iteration on the 2D Poisson equation in a structured domain are developed by Venkatasubramanian and Vuduc. Their best implementation is able to achieve $98\%$ GPU bandwidth. To further improve performance, they develop asynchronous implementations which remove the need for complete synchronization each step, leading to a 1.2 to 2.5 times speedup over their previously optimal implementation \cite{Vuduc2009}. Ahamed and Magoul\'es develop a Krylov-like version of Jacobi iteration and apply it to several three dimensional problems (Laplace equation, gravitational potential equation, heat equation) on the GPU, observing a 23 times speedup using this new version compared to the serial CPU-based algorithm \cite{Ahamed2016}.

The studies listed above detail the application of Jacobi iteration on the GPU to a variety of problems, such as 2D structured problems and full 3D unstructured problems. In all cases, the studies highlight the need for efficient memory access for optimal performance. Memory bandwidth, not computational complexity, has emerged as the limiting factor in scientific simulations. Therefore, it is essential that applications utilizing the GPU respect its memory hierarchy in an effort to achieve the best performance.

In this work, we develop an algorithm to perform the Jacobi iterative method efficiently using on-chip GPU shared memory, which is quicker to access than GPU global memory. We focus specifically on structured matrices. Shared memory is accessible on the GPU as a cache-like memory which has much higher bandwidth compared to global memory which is traditionally used on the GPU \cite{Hwu2016}. However, shared memory is typically very small (16 kB - 48 kB based on settings and device compute capability \cite{Harris2013}) and only exists per GPU block. This inspires a domain decomposition style algorithm where the domain is separated into subdomains which are allocated to the shared memory of different GPU blocks. Within the blocks, we perform many iterations of Jacobi, avoiding communication between GPU blocks at every step. This results in a hierarchical algorithm in which every cycle, multiple Jacobi iterations are performed independently within the shared memory of each GPU block (termed subiterations) and copied back to global memory (which holds the most up to date solution at the end of the cycle). Previous studies have generally focused on the performance of the standard Jacobi iteration using shared memory. However, this hierarchical approach results in a variation of Jacobi iteration that demonstrates better performance in parallel settings where a memory hierarchy exists (such as on the GPU). We report comparisons between a classical GPU implementation of Jacobi iteration as well as this shared memory approach. 

The rest of our paper is structured as follows. Section 2 provides background for this work; a review of the Jacobi iterative method for solving linear systems of equations and an overview of the CUDA framework for GPU computing. Section 3 introduces various approaches for performing Jacobi iteration, including our new shared memory approach. and provides performance comparisons between this approach and a traditional GPU approach with global memory for the 1D Poisson equation. Section 4 presents a similar algorithm and comparison results in the context of 2D problems. Lastly, Section 5 provides concluding remarks on this work.

\section{Background}

\subsection{Jacobi Iterative Method}

Jacobi iteration is an iterative method used to solve a linear system $Ax = b$ where $A \in \mathbb{R}^{n \times n}$ and $x \in \mathbb{R}^n$, $b \in \mathbb{R}^n$. The Jacobi iterative method is an example of a stationary iterative method which involves a matrix splitting of the form $A = M - N$ \cite{ParallelIterativeAlgorithms2007}. Then the iterative update is given by:
\begin{equation}
    x^{(n+1)} = M^{-1} (b - N x^{(n)})
\end{equation}
For Jacobi iteration, $M \equiv D$ where $D$ is a diagonal matrix with diagonal entries equal to those of the matrix $A$. Meanwhile, the matrix $N \equiv A - D$. Elemental Jacobi iteration can be written as
\begin{equation}
    x_i^{(n+1)} = \frac{1}{a_{ii}} \left(b_i - \sum_{j \neq i} a_{ij} x_j^{(n)} \right)
    \label{eq:jacobi-elemental}
\end{equation}
In order for Jacobi iteration to converge, the spectral radius of the iteration matrix must be less than one
\begin{equation}
    \rho(D^{-1} (A-D)) < 1
\end{equation}
Jacobi iteration can be used to solve linear systems which involve matrices that are diagonally dominant. Many matrices arising from discretizations of PDEs are diagonally dominant and can be solved using Jacobi iteration.

\subsection{CUDA Framework for GPUs}

A schematic of the CUDA model is shown in Figure 1. 
\begin{figure}[htbp!]
\centering
\includegraphics[width=0.4\textwidth]{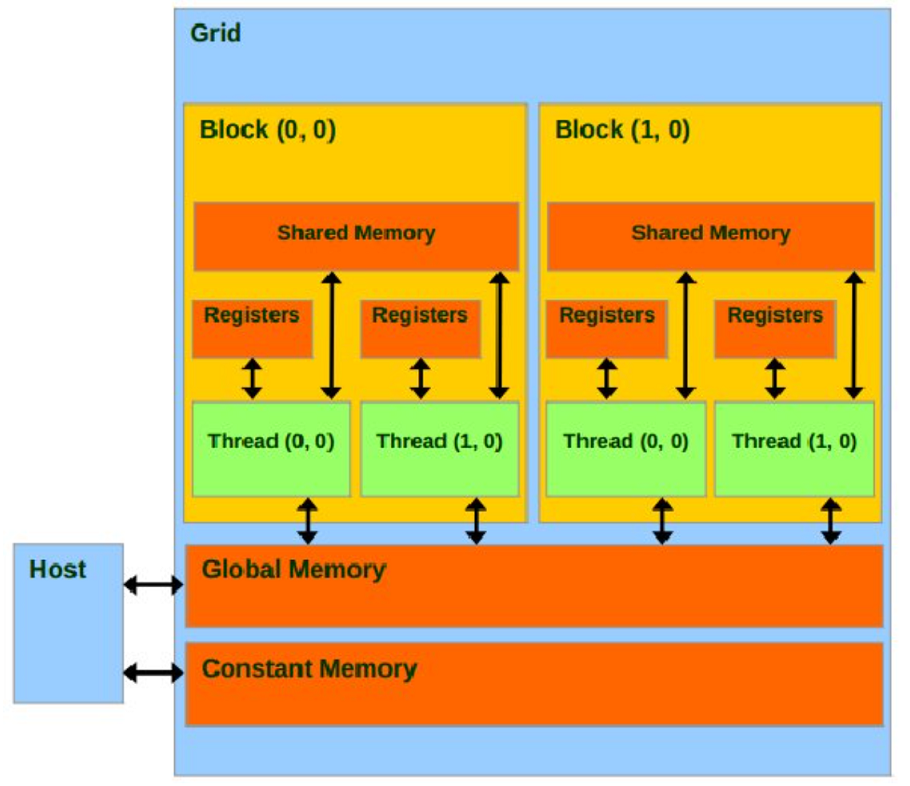}
\caption{CUDA Hierarchy and Memory Model (taken from \cite{Hwu2016}).}
\end{figure}
A GPU program developed using CUDA (the parallel computing platform developed by NVIDIA for general purpose GPU programming) typically begins with initialization on the CPU (termed host) and transfers to the GPU (termed device) where parallel calculations will be performed. The most fundamental unit of parallel execution is the thread. A large number of threads can run simultaneously. This gives GPUs high computing power and the ability to perform mathematical operations much more quickly than CPUs. Groups of threads are organized into blocks (up to 1024 threads per block are permitted on most modern machines). Every thread is assigned a thread ID that is unique to it in the thread block, given by $\texttt{threadIdx.x}$. For 2D and 3D problems, a $\texttt{threadIdx.y}$ and $\texttt{threadIdx.z}$ are also assigned (these are set to 0 by default for the 1D case). Similarly, every block has its own unique block ID given by $\texttt{blockIdx.x}$ (a similar $\texttt{blockIdx.y}$ and $\texttt{blockIdx.z}$ are assigned in 2D and 3D). Each block contains its own shared memory storage which is accessible to the threads comprising the block. 

When launching a kernel (calculation to be performed on the device), the user specifies the number of threads per block (denoted by $\texttt{blockDim.x}$; $\texttt{blockDim.y}$ and $\texttt{blockDim.z}$ are also defined for 2D and 3D problems) as well as the total number of blocks to be invoked. The blocks are then distributed between streaming multiprocessors (SMs), and the calculations are launched by warps (groups of 32 threads belonging to the same block). It is preferable to select a thread per block value which is a multiple of 32 so that all threads in a warp are utilized. The warps are scheduled in a round-robin fashion so that latency is hidden as calculations are being performed. They continue to operate until all threads have completed their calculations. 

When a block is allocated to an SM, the block receives a portion of the total shared memory on the SM. The amount of shared memory in a block is usually limited (modern GPUs typically have 48 kB of shared memory per block) and accessible only to threads within a block. However, shared memory is more quickly accessible (approximately 10 times higher bandwidth \cite{Wilt2013} and 100 times lower latency \cite{Harris2013}) than the GPU global memory. Therefore, efficient use of intra-block shared memory can reduce runtime significantly.

\section{Approaches for Jacobi Iteration in 1D Problems}

We introduce a Jacobi iterative solver which utilizes GPU shared memory for one dimensional problems. In general, linear systems arising from one dimensional partial differential equations (PDEs) can be solved efficiently using direct methods (e.g. Thomas algorithm for tridiagonal systems), with iterative methods being desirable when dealing with linear systems arising from discretizations of higher dimensional PDEs. However, applying our shared memory Jacobi iterative solver to the 1D case can provide insights which translate to higher dimensions.

\begin{figure}[htbp!]
    \centering
    \includegraphics[width=0.7\textwidth]{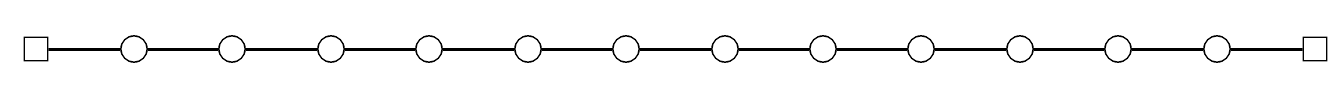}
    \caption{A sample one-dimensional domain. Square points indicate Dirichlet boundary conditions.}
    \label{fig:1D-domain}
\end{figure}

\begin{figure}[htbp!]
\centering
\includegraphics[width=0.2\textwidth]{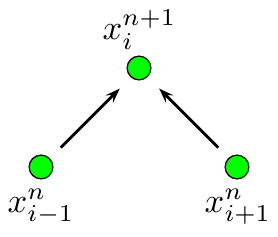}
\caption{Stencil for Jacobi iteration performed on a triadiagonal matrix. The value of a degree of freedom at the next step depends only on adjacent values at the current step.}
\label{fig:stencil}
\end{figure}

We consider a linear system which arises from a 1D PDE (e.g. diffusion equation) on a grid as shown as in Figure \ref{fig:1D-domain}. The interior points correspond to the different degrees of freedom (DOFs) of a linear system. We assume Dirichlet BCs, represented by the square endpoints.
One-dimensional PDEs discretized with standard discretizations (finite difference, finite elements) usually result in a tridiagonal system. Consider a general tridiagonal matrix $A$ with the values $a_{i}$, $d_{i}$, and $c_{i}$ in the $i$th subdiagonal, diagonal, and superdiagonal entries, respectively (as shown below). 
\[
A =
\begin{pmatrix}
d_1 & c_1 &  \\
a_2 & d_2 & c_2 \\
 & \ddots & \ddots & \ddots \\
 &  & a_{n-1} & d_{n-1} & c_{n-1} \\
 &  &  & a_{n} & d_{n}
\end{pmatrix}
\] 
Applying the Jacobi iterative method (Equation \eqref{eq:jacobi-elemental}) to advance the $i$th DOF (denoted by $x_i$) in our tridiagonal system from iteration $n$ to iteration $n+1$ results in the following update equation (where $b_i$ represents the right hand side value at the $i$th grid point)
\begin{equation}
x_{i}^{n+1} = \frac{1}{d_{i}}(b_{i} - a_{i} x_{i-1}^{n} - c_{i} x_{i+1}^{n})
\label{eq:jacobi-stencil-equation-1d}
\end{equation} 
For a particular DOF, Equation \eqref{eq:jacobi-stencil-equation-1d} shows that the solution at the next iteration depends only on the values at adjacent DOFs at the current iteration. This is illustrated in Figure \ref{fig:stencil}. If the neighbor values are available, we can perform the update on the DOF. This idea is used to develop our various approaches for solving the 1D problem on the CPU and GPU.

\subsection{CPU Approach}
\label{sec:1d-cpu-algorithm}
Jacobi iteration can be implemented on the CPU as follows. Each degree of freedom in the grid can be updated in a sequence, with the subsequent values at step $n+1$ relying only on the current values at step $n$. This requires two storage containers, one to store the solution at step $n$ and the other to store the solution at step $n+1$. Once all DOFs are updated in a given step, the two arrays can be swapped and the process can be repeated. A sample C++ style code to illustrate the CPU approach is shown below. The $\texttt{jacobiUpdate}$ function corresponds to the update equation in Equation \eqref{eq:jacobi-stencil-equation-1d}.

\lstset{language=C++}  
\begin{lstlisting}
Given x0, b:
double * x1 = new double[N];

for (int k = 0; k < numSteps; k++) {
    for (int i = 1; i < N-1; i++) {
        double left_value = x0[i-1];
        double right_value = x0[i+1];
        double rhs_value = b[i];
        x1[i] = jacobiUpdate(left_value, right_value, rhs_value);
    }
    swap(x0, x1);
}
\end{lstlisting}

\subsection{GPU Approach with Global Memory}
\label{sec:1d-gpu-algorithm}
The GPU approach for implementing Jacobi iteration with global memory is quite similar to the CPU approach. Rather than use a for loop, each GPU thread can independently update each DOF. We must ensure that we request enough threads to perform the updates on all DOFs. We can do this by setting any value for the number of threads per block (preferably a multiple of 32 as threads operate in warps containing 32 threads), and setting the number of blocks equal to the ceiling of the total number of DOFs divided by the number of threads per block specified earlier. A sample CUDA C++ style code to illustrate the GPU approach is shown below.

\begin{lstlisting}
Given x0, b:
double * x1 = new double[N];

for (int k = 0; k < numSteps; k++) {
    int i = threadIdx.x + blockDim.x * blockIdx.x;
    if (i > 0 && i < N-1) {
        double left_value = x0[i-1];
        double right_value = x0[i+1];
        double rhs_value = b[i];
        x1[i] = jacobiUpdate(left_value, right_value, rhs_value);
    }
    __syncthreads();
    swap(x0, x1);
}
\end{lstlisting}

\subsection{GPU Approach with Shared Memory}
\label{sec:1d-shared-algorithm}
Shared memory on the GPU has a higher bandwidth and lower latency compared to global memory \cite{Wilt2013, Harris2013}, so adapting GPU applications to use shared memory has become a popular optimization technique for improving performance. Performing Jacobi updates within shared memory can improve performance because reading current values to and writing updated values from shared memory will be much faster compared to performing these accesses in global memory. However, using shared memory can be challenging because it is allocated on a per block basis. Threads in one GPU block cannot access the shared memory of another GPU block. This is a byproduct of on-die shared memory being very limited. Only a set number of blocks can be active at a given time and receive a shared memory allocation. In general, blocks must wait in the queue to operate, as well as receive a shared memory allocation and perform updates. Although the limited availability of shared memory makes it challenging to use, it can provide speedup in many applications which are memory bound, such as iterative solvers.

\begin{figure}[htbp!]
    \centering
    \includegraphics[width=0.9\textwidth]{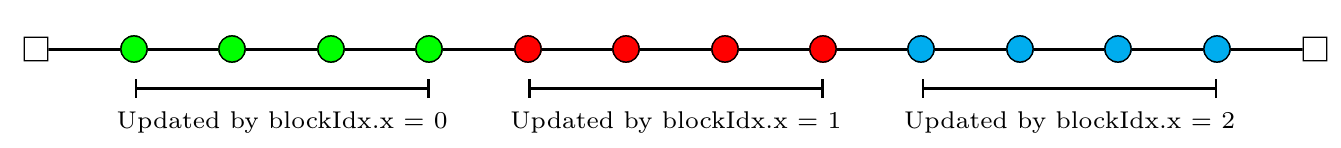}
    \caption{1D domain partitioned between different GPU blocks, Each GPU block is responsible for updating its own set of points.}
    \label{fig:partitioned-domain}
\end{figure}

One way to utilize shared memory in this application is to partition the solution data between the available shared memory and operate on these partitions independently. For the 1D problem, the DOFs in the domain shown in Figure \ref{fig:1D-domain} can be partitioned into several subdomains. The DOFs in each subdomain may be updated by a distinct GPU block. Figure \ref{fig:partitioned-domain} shows an example of a 12 point grid, divided into 4 point subdomains which will be operated on by 3 GPU blocks. 

\begin{figure}[htbp!]
    \centering
    \includegraphics[width=0.9\textwidth]{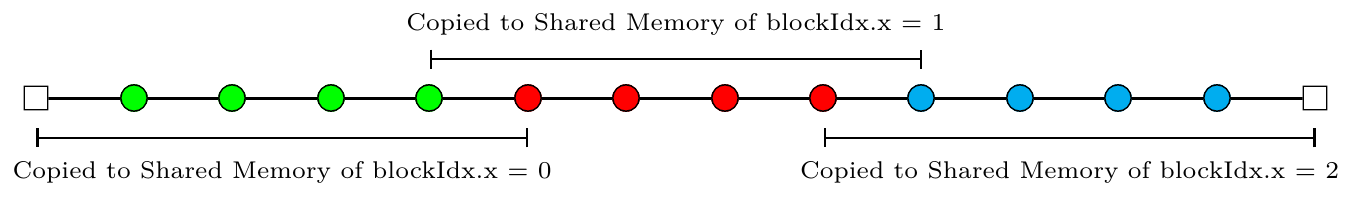}
    \caption{Copy of data from global memory (containing the entire grid) to the shared memory of each GPU block. Each block receives data corresponding to points it is responsible for updating plus an additional left and right point.} 
    \label{fig:allocated-grid}
\end{figure}

The first step in our approach is to copy the DOF values in each subdomain from global memory (where the initial solution resides) to the shared memory of the GPU block which will be responsible for its update. Updating the edge values in each subdomain requires information from the adjacent DOFs which will not be accessible to the block. We address this by augmenting the subdomains by an extra point to the left and right (beyond the points that will be updated). These augmented subdomains (shown in Figure \ref{fig:allocated-grid} for our example) can be copied to the shared memory of each GPU block. The shared memory simply holds a temporary copy of the solution which is to be operated on, while the final solution is always stored in global memory. 

After the copy to shared memory is complete, we can perform Jacobi iteration on the interior points of each subdomain using the stencil update given in Equation \eqref{eq:jacobi-stencil-equation-1d}. To maximize parallelism, each DOF update should be handled by a distinct thread. For our example, the number of threads per block should be set to 4 so that 4 threads can be used to update the 4 interior points within each subdomain. In general, the threads per block specification is exactly equal to the number of interior points in a subdomain (i.e. the subdomain size is directly tied to the user specified thread per block choice).  Given a thread per block specification of $\texttt{blockDim.x}$, the subdomain size will be $\texttt{blockDim.x + 2}$ with  $\texttt{blockDim.x}$ interior points to be updated. To execute the previous global to shared memory transfer efficiently, each of the $\texttt{blockDim.x}$ threads should copy one value from global to shared memory and the remaining two values can be copied by threads 0 and 1. One can alter the subdomain size by changing the thread per block specification. For efficiency, the threads per block specification should be a multiple of 32 because threads are launched in groups of 32 threads (known as warps). This ensures all threads in a warp are active and avoids thread divergence. As a consequence, subdomains with a multiple of 32 interior DOFs are preferable.

Since data movement tends to be the bottleneck in iterative linear solvers (rather than computation), it is beneficial to perform many updates within shared memory before executing expensive data transfers back to global memory. Following this logic, we recommend performing many Jacobi iterations within shared memory. We refer to the Jacobi iterations performed within shared memory as subiterations, and denote the number of subiterations performed by $k$. Upon performing the $k$ subiterations, the updated values can be copied from the shared memory of every GPU block to their global memory position. Only the interior $\texttt{blockDim.x}$ values of each subdomain must be copied back from shared to global memory since the two edgemost points are not updated. Each thread should copy the same DOF it updated from shared to global memory. Shared memory is terminated at the end of this step and the updated solution now exists in global memory.

\begin{figure}[htbp!]
    \centering
    \includegraphics[width=\textwidth]{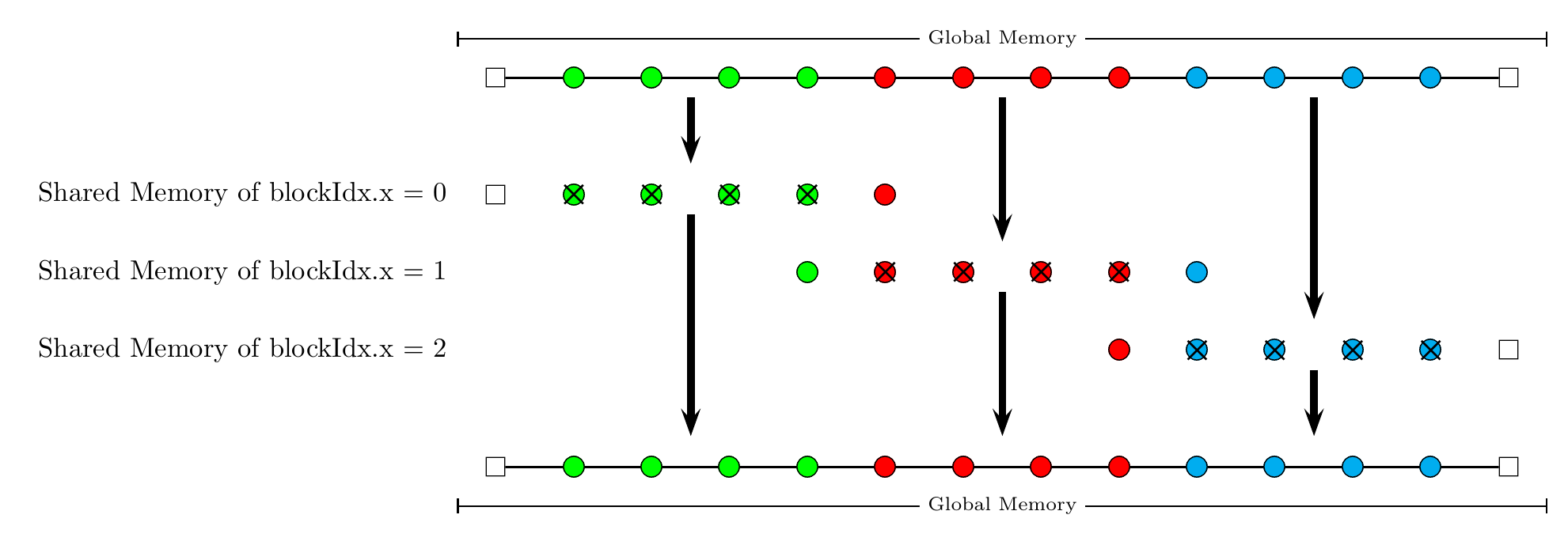}
    \caption{Schematic of the shared memory algorithm for 1D Jacobi iteration. Initially, data is copied from global memory to the shared memory of different GPU blocks. Each GPU block is responsible for updating a distinct set of points. After performing $k$ Jacobi iterative updates (marked by X) the updated point values are transferred back to global memory.}
    \label{fig:shared-to-global-memory-transfer}
\end{figure}

This completes one cycle of our shared memory algorithm. The algorithm is implemented in a single GPU kernel (since shared memory is only active as long as a GPU kernel is running). A sample CUDA C++ style code to illustrate the implementation is provided in Appendix A. The main parameters in the algorithm are the number of threads per block ($\texttt{blockDim.x}$) and the number of subiterations ($k$). The number of threads per block controls the size of the subdomains in shared memory, while the number of subiterations controls the number of Jacobi iterations performed within shared memory before communication back to global memory. Our algorithm is depicted in Figure \ref{fig:shared-to-global-memory-transfer}. In summary, each cycle of the algorithm involves the following three steps:
\begin{enumerate}
\item Partition the gridpoints between the different GPU blocks and copy values from global memory to shared memory. Each subdomain size has size $\texttt{blockDim.x} + 2$.
\item Update the interior $\texttt{blockDim.x}$ points in each GPU block by performing $k$ Jacobi iterations (we refer to these as subiterations). 
\item Update the global memory solution array by copying the interior $\texttt{blockDim.x}$ values from shared memory to their appropriate positions in global memory (shared memory is terminated at this point).
\end{enumerate}

An application that uses GPU shared memory requires the user to specify the number of bytes of shared memory needed. At the very least, there must be enough space for twice the subdomain data (which has size $\texttt{blockDim.x + 2}$) because Jacobi iteration requires two storage containers (one containing the current solution values and one containing the updated values). During even subiterations, the first \texttt{blockDim.x + 2} entries within shared memory should hold the current solution values while the updated values are placed in the second \texttt{blockDim.x + 2} entries. During odd subiterations, this should be reversed with the second \texttt{blockDim.x + 2} entries holding the current solution values while the updated values are placed in the first \texttt{blockDim.x + 2} entries. The $\texttt{blockDim.x}$ right hand side values corresponding to the interior subdomain DOFs can also be copied to shared memory (as the right hand side value appears in the Jacobi update function in Equation \eqref{eq:jacobi-stencil-equation-1d}). Therefore, the total amount of shared memory required for the algorithm is 2 times the subdomain size ($\texttt{blockDim.x + 2}$) for the solution array plus an array of size $\texttt{blockDim.x}$ for the right hand side values corresponding to the interior DOFs. This should be multiplied by the size of the representation (4 bytes for floats, or 8 bytes for doubles).

One practical question to consider when applying the algorithm is what is a good number of subiterations to perform within shared memory.  Using a large value of $k$ is beneficial as we can perform many Jacobi updates without requiring expensive global memory accesses. The drawback of setting a large $k$ is the lack of frequent update of the boundary values of each subdomain. These boundary points are updated by the adjacent blocks, and the most up to date values are available to a subdomain only after each cycle when all values are transferred back to global memory after the subiterations are performed and reallocated back to shared memory. Therefore, subiterations used to update interior points within a subdomain rely on old boundary data, which can negatively impact convergence. Setting $k = 1$ ensures that subdomains have the most up to date boundary values at every step, and is exactly equivalent to performing Jacobi iteration. However, setting such a low $k$ value requires many expensive global memory accesses which will inhibit algorithm performance. There is a tradeoff between receiving up to date subdomain values more frequently to improve convergence (by setting a small $k$), and performing fewer global memory accesses to improve performance (by setting a small $k$). We explore good values for $k$ in our numerical tests.

\subsection{Effect of Subiterations}
The classic GPU and shared memory approaches for Jacobi iteration presented in Sections \ref{sec:1d-gpu-algorithm}-\ref{sec:1d-shared-algorithm} are applied to the triadiagonal system arising from a finite difference discretization of the Poisson equation on a one dimensional domain $x \in [0, 1]$ with Dirichlet boundary conditions (shown in Equation \eqref{eqn:poisson}):
\begin{equation}
-\frac{d^2 u}{dx^2} = f, \ \ u(0) = u(1) = 0
\label{eqn:poisson}
\end{equation}
The triadiagonal system is given by
\begin{equation}
A x = b
\label{fig:linear-system}
\end{equation}
where 
\[
A = \frac{1}{\Delta x^2}
\begin{pmatrix}
2 & -1 &  \\
-1 & 2 & -1 \\
 & \ddots & \ddots & \ddots \\
 &  & -1 & 2 & -1 \\
 &  &  & -1 & 2
\end{pmatrix}, \ \
b = \begin{pmatrix}
1   \\
\vdots \\
\vdots \\
\vdots \\
1 \\
\end{pmatrix}
\] 
The right hand side vector $b$ has all entries set to one.  We consider a problem size of $N = 1024$ interior DOFs (1026 points in total) and measure the time required to reduce the $L_{2}$ residual norm of the initial solution (set to a vector of ones) by a factor of 1e-4 using Jacobi iteration. The elemental Jacobi update (Equation \eqref{eq:jacobi-elemental}) applied to the 1D Poisson equation results in the update equation \eqref{eq:jacobi-stencil-1d-poisson}.
\begin{equation}
    x_i^{(n+1)} = \frac{b_i (\Delta x)^2 + x_{i-1} + x_{i+1}}{2}
    \label{eq:jacobi-stencil-1d-poisson}
\end{equation}
Because this problem is too small to fully utilize the streaming multiprocessors of the GPU, $1024$ copies of the problem are operated on simultaneously. This ensures that the GPU is fully utilized and that our algorithm can scale to higher dimensional problems (i.e. an analogous 2D problem on a 1024 by 1024 grid). The time required for the classic GPU and the shared memory approaches to decrease the residual for the 1024 copies is measured.  Our main goal is to understand the speedup we can achieve using shared memory over a classical GPU implementation of Jacobi iteration. All numerical tests were run using a TITAN V GPU with double precision.

The thread per block specification is varied between the values of [32, 64, 128, 256, 512] for the classic GPU approach, and the best time is recorded. For the shared memory implementation, the thread per block value of 32 is used. This corresponds to a required shared memory allocation of 800 bytes per block (much lower than the shared memory limit). Although larger thread per block values could be used in 1D, we wish to extend our findings to higher dimensions in which case the available threads would have to be split between different dimensions. The number of blocks is always set such that there are enough to accommodate all of the DOFs in the problem (i.e. ceiling of the total number of DOFs from all 1024 copies of the problem divided by the thread per block value). The following 6 subiteration values are explored in this study: $k = 4, 8, 16, 32, 64, 128$.

\begin{figure}[htbp!]
    \centering
    \includegraphics[width=0.6\textwidth]{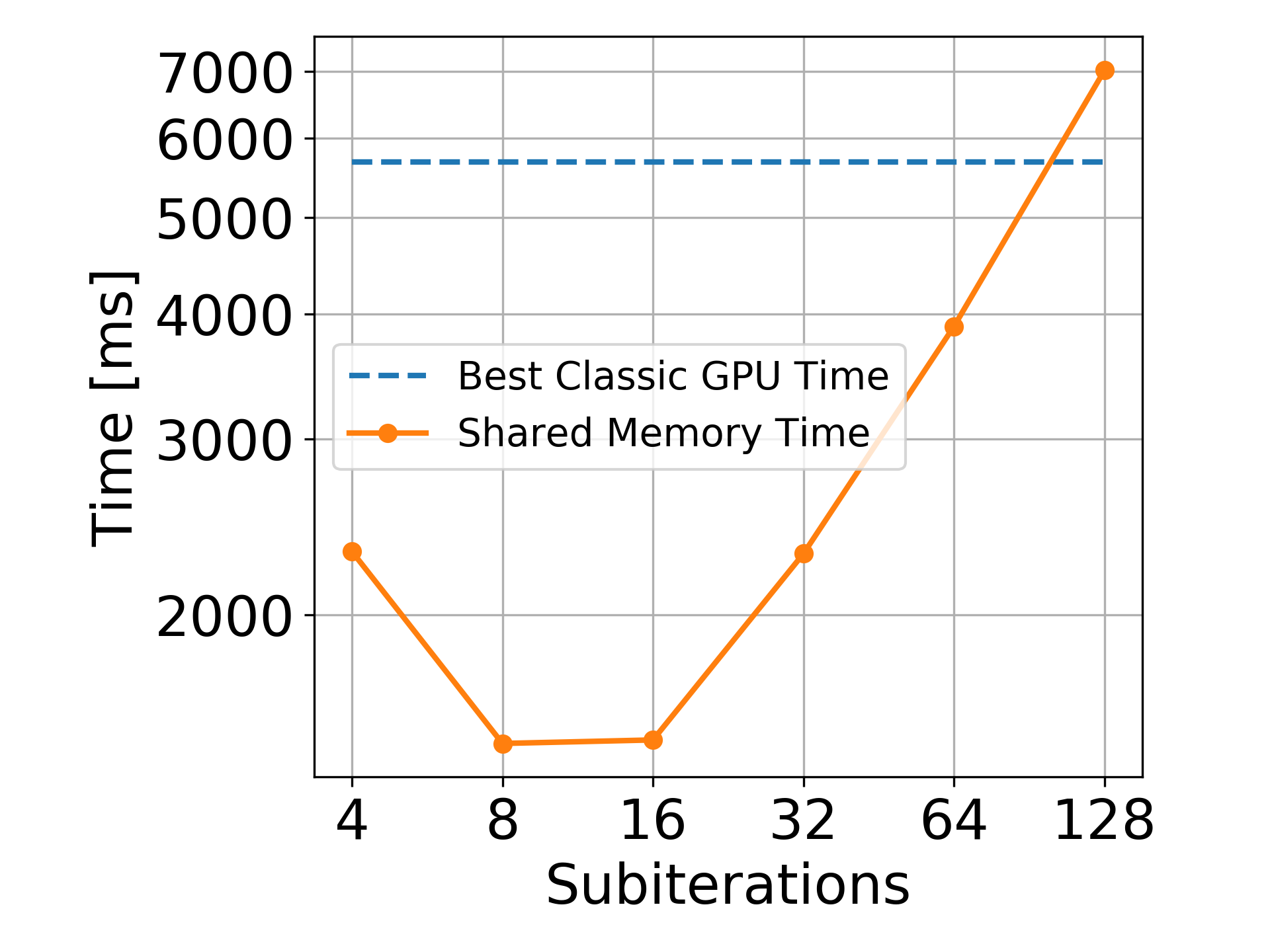}
    \caption{Comparison of time required for classic GPU and shared memory implementations with different subiteration values to reduce the residual of the solution by 1e-4 for a problem size of $N = 1024$ (solving 1024 copies simultaneously). The shared memory approach outperforms the classic approach in all cases except when $k = 128$, and performs the best when $k = 8$.}
    \label{fig:gpu-vs-shared}
\end{figure}

Figure \ref{fig:gpu-vs-shared} shows the performance of the shared memory approach for the various subiteration values specified above compared to the best classic GPU performance. In each case, the number of iterations/cycles required for convergence was determined ahead of time, and the GPU and shared memory implementations were run under the prescribed number of iterations/cycles to avoid calculating the residual during these time trials. The performance results include the time required for host to device and device to host transfers, and were recorded using the CUDA Event API. For the thread per block values of [32, 64, 128, 256, 512], the classic GPU times measured were [7248.11, 5690.35, 5675.69, 5677.27, 5680.42] ms. The optimal time measured is 5675.69 ms when the thread per block value is 128, although the time is essentially constant for thread per block values of 64 and above. In these cases, the GPU is being fully utilized so the performance cannot be further improved. The shared memory approach outperforms the classic GPU approach in all cases except when the number of subiterations is set to the highest value of 128. The best performance is obtained when $k = 8$. Past $k = 16$, the time to converge increases linearly with the number of subiterations, suggesting that performing too many subiterations within shared memory can hurt performance.

Table \ref{tab:gpu-vs-shared-speedup} shows the speedup achieved going from the best classic GPU performance to the shared memory performance for each subiteration value specified. The best speedup achieved is nearly four times over the classic GPU approach when $k = 8$ (setting $k = 16$ results in a nearly similar speedup). The results demonstrate that the time for convergence for Jacobi iteration can be reduced using shared memory. 

\begin{table}[htbp!]
    \centering
    \begin{tabular}{|c|c|c|c|c|c|c|}
    \hline
    Number of Subiterations & 4 & \textbf{8} & 16 & 32 & 64 & 128 \\
    \hline
    Best GPU to Shared Speedup & 
    2.45 & \textbf{3.82} & 3.79 & 2.47 & 1.46 & 0.81 \\
    \hline
    \end{tabular}
    \caption{Speedup from classic GPU to shared memory approach for various subiteration values. A nearly four times speedup is achieved when $k = 8$.}
    \label{tab:gpu-vs-shared-speedup}
\end{table}

\subsection{Effect of Overlapping Subdomains in 1D}
\label{sec:overlap}
One key feature of our shared memory algorithm is that many Jacobi updates are performed each cycle without transfer to global memory. While this minimizes global memory accesses to improve performance, convergence is hampered by the lack of up to date values used for the updates. In particular, boundary data in each subdomain is only updated every cycle (after $k$ subiterations), so many cycles may be required for convergence. DOFs close to the subdomain edges are especially affected and contribute a larger residual value to the overall residual norm. Improving the edge values can greatly improve the convergence of the algorithm and reduce the number of cycles necessary. 

We attempt to alleviate this problem by using overlapping subdomains. The key idea of this approach is that DOFs which are further away from subdomain edges (closer to the subdomain center) have less error than those closer to the edges. With overlapping subdomains, we permit sets of points close to the subdomain edges to be allocated to two subdomains (copied to the shared memory of two GPU blocks). A DOF which exists in two subdomains will have a different local position in each subdomain. The subdomain in which the DOF is further from the edge is the one which will contribute the final value to global memory.

\begin{figure}[htbp!]
    \centering
    \includegraphics[width=\textwidth]{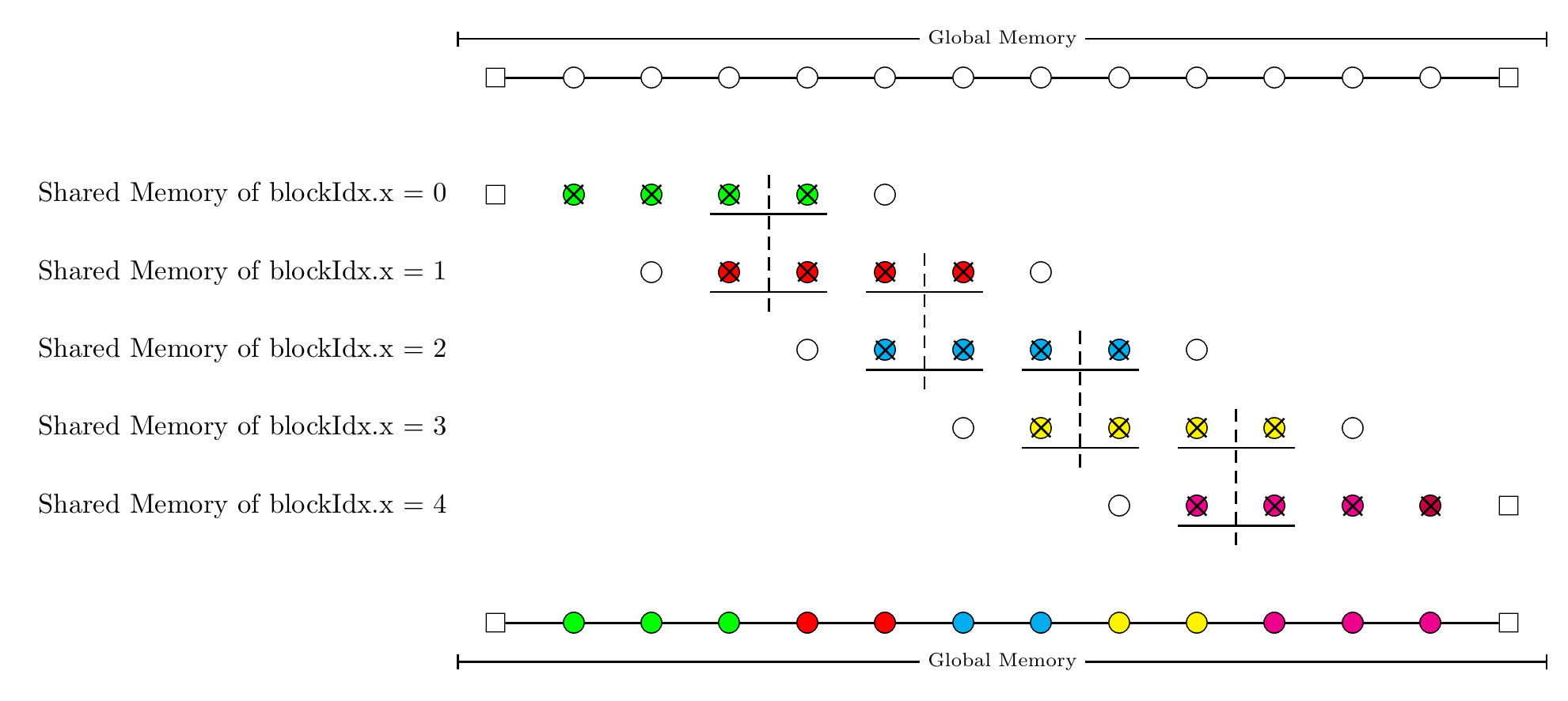}
    \caption{Schematic of the transfer of data between global memory array and shared memory of GPU blocks, when subdomains have overlapping points (in this example, two interior points overlap between each subdomain). This approach improves the global memory solution of interior DOFs which reside near subdomain edges in shared memory.}
    \label{fig:1d-overlapping-approach}
\end{figure}

To illustrate this approach more clearly, an example is shown in Figure \ref{fig:shared-to-global-memory-transfer} for the 12 point grid. In this example, each subdomain has two interior DOFs in common with its neighbor. These overlapping DOFs are updated in the shared memory of two GPU blocks. Although either block could copy its DOF value back to global memory at the end of the cycle, we choose to have the left subdomain provide the final left point value and the right subdomain provide the final right point value in each set of overlapping points. This is because the local position of the left point within the left subdomain is further from the edge while compared to its local position in the right subdomain. Meanwhile, the local position of the right point is further from the edge in the right subdomain than in the left subdomain. Figure \ref{fig:1d-overlapping-approach} shows how the GPU blocks contribute to the overall global memory solution in this approach. In general, any even number of overlapping points is permitted (as long as it is less than the number of interior points in the subdomain) and in this case the left subdomain should contribute the left half of points while the right subdomain should contribute the right half.

Overlapping points between different subdomains and selectively copying DOF values as described will reduce the number of cycles required for convergence. This improvement in convergence is similar to behavior seen in overlapping domain decomposition methods, where the convergence rate is improved as more overlap is introduced between subdomains \cite{Dolean2015}. A drawback of overlapping subdomains is that multiple GPU blocks must perform updates on many of the points throughout the domain. This requires utilization of more GPU blocks, as the overlap effectively reduces the number of distinct points each subdomain handles. The number of overlapping points is another algorithm parameter which must be chosen judiciously. 

\begin{figure}[htbp!]
    \centering
    \begin{subfigure}[b]{0.6\textwidth}
        \includegraphics[width=\textwidth]{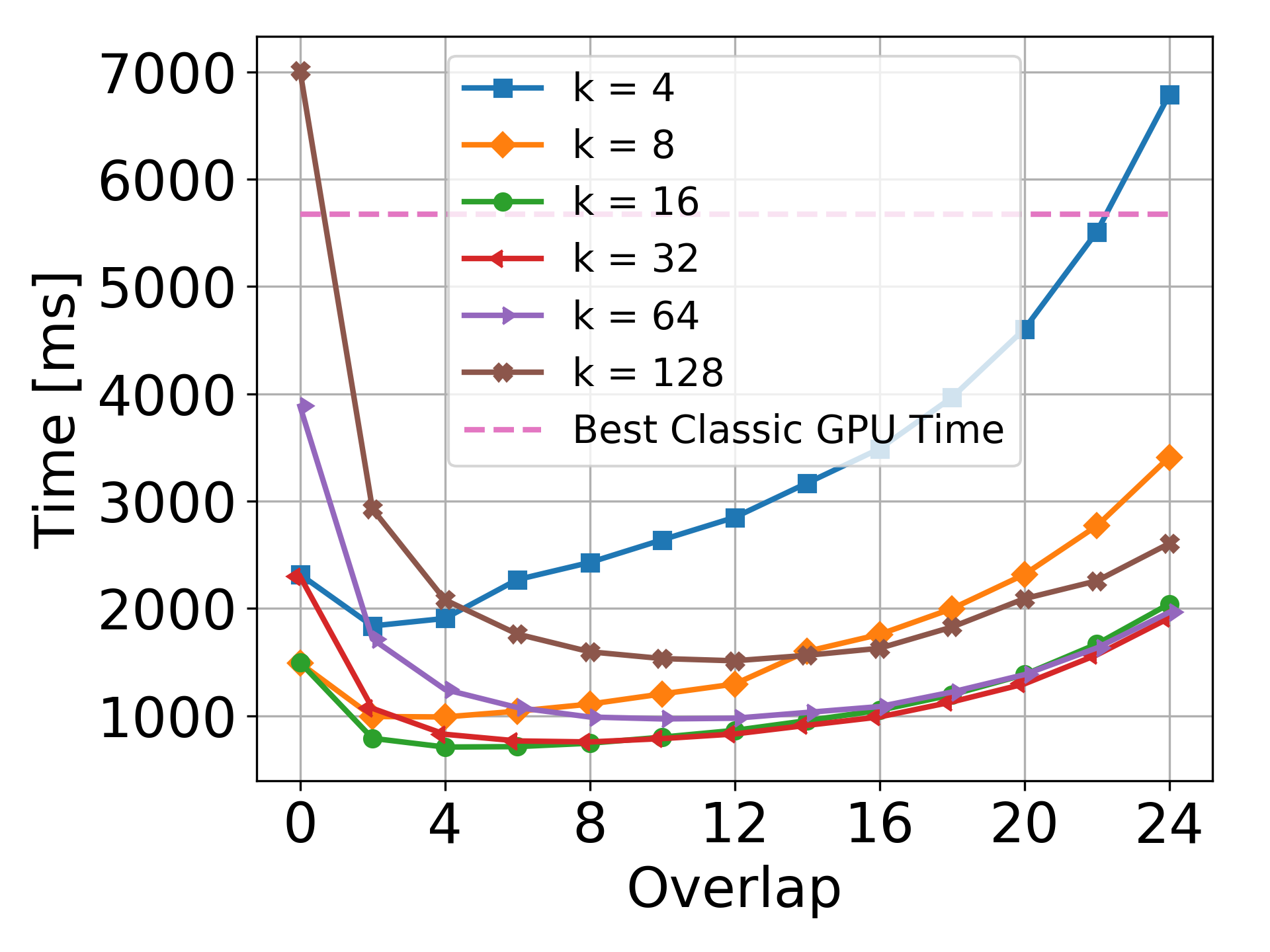}
        \caption{Time}
         \label{fig:subiteration-time}
    \end{subfigure}
    \hfill
    \begin{subfigure}[b]{0.6\textwidth}
        \includegraphics[width=\textwidth]{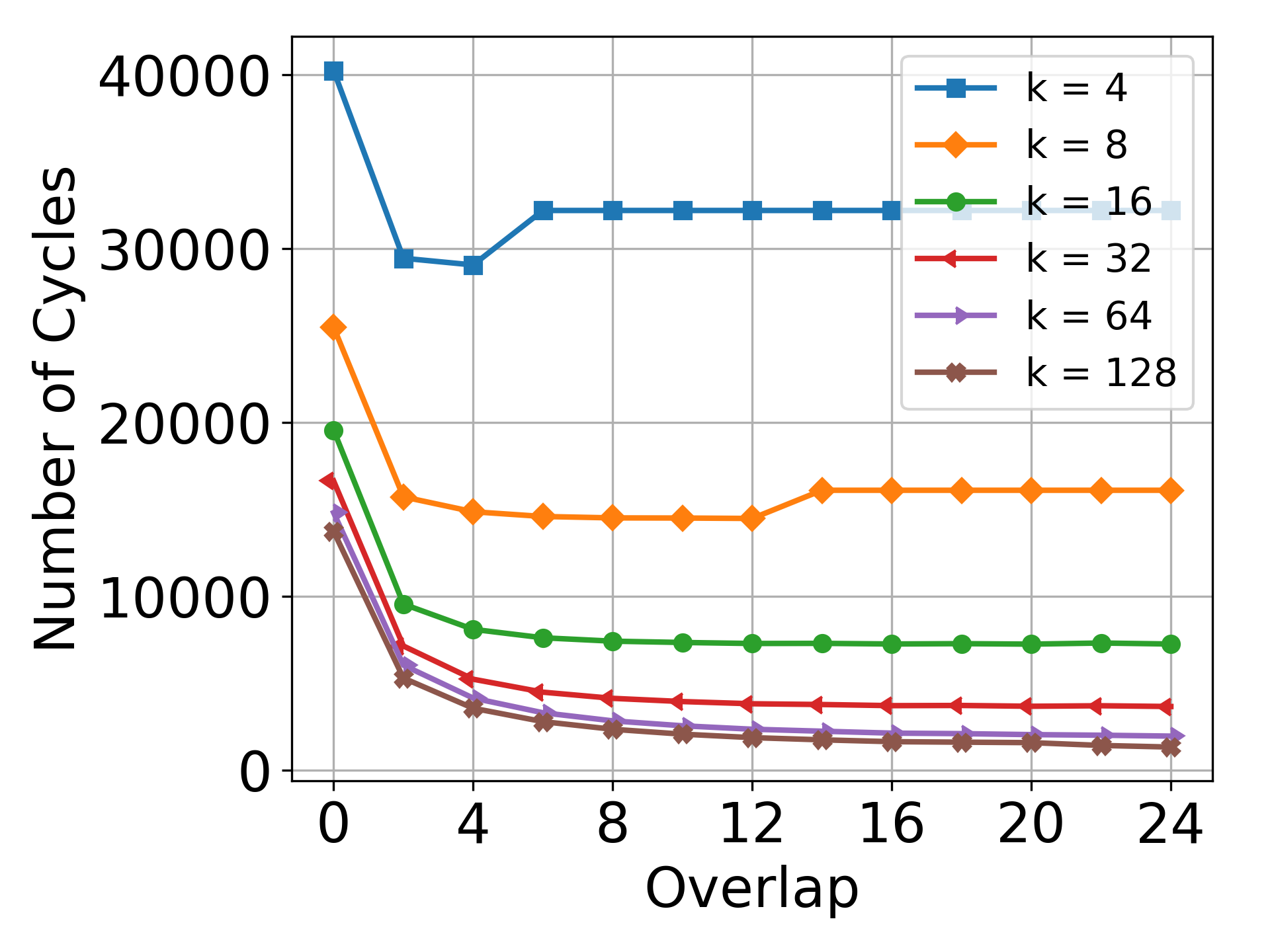}
        \caption{Number of Cycles}
        \label{fig:subiteration-cycles}
    \end{subfigure}
    \caption{Time and Number of Cycles required to reduce the residual as a function of overlap for different subiterations $k$. For a given $k$, performing overlap initially decreases the number of cycles required for convergence (causing the time to decrease as well). The number of cycles eventually saturates as overlap increases further. However, the time increases due to the extra computational resources required. The best performance is achieved with $k = 16$.}
\end{figure}

Numerical tests are performed to study the effect of overlap on the performance of the shared memory algorithm. In these tests, a thread per block value of 32 is used as before. Each subdomain can have an even number of interior overlapping points with its neighboring subdomain (as long as the number of overlapping points is less than the number of interior points in a subdomain). For our thread per block specification of 32, we permit up to 30 interior overlapping points between subdomains. The left subdomain will contribute the left half of the overlapping points while the right subdomain contributes the right half. Figures \ref{fig:subiteration-time} and \ref{fig:subiteration-cycles} show the time and number of cycles required for convergence as a function of overlap for subiteration values of $k = 4, 8, 16, 32, 64, 128$. 

For all subiteration cases, allowing just two points to overlap causes a drop in the number of cycles required for convergence. This is a result of improving poor DOF values adjacent to subdomain edges. The reduction in cycles leads to an initial reduction in time, particularly for larger values of $k$. As the amount of overlap increases further the number of cycles required for convergence continues to decrease slowly and eventually saturates. The time; however, grows dramatically as overlap is increased further. This is because the amount of computation required per cycle grows with overlap. Given a problem size with $N$ interior DOFs, threads per block $tpb$ and overlap $o$, the number of operational blocks required is:
\begin{equation}
    \text{Operational blocks} = \frac{N - tpb}{tpb - o} + 1
\end{equation}
The logic for this is as follows. There are initially $N+2$ points in the grid (including boundary points). The first block has a subdomain size of $tpb + 2$ points, and every subsequent block added handles $tpb - o$ unique points in our grid. Therefore, subtracting the number of points handled by the initial block from $N+2$ and dividing by the number of points handled by subsequent blocks gives the number of blocks required. This can also be written as:
\begin{equation}
    \text{Operational blocks} = \frac{N - tpb}{tpb - o} + 1 = \frac{N - tpb + tpb - o}{tpb - o} = \frac{N - o}{tpb - o}
\end{equation}
The corresponding number of threads required is the number of blocks times the threads per block value $tpb$.
\begin{equation}
    \text{Operational threads} = (\text{Operational blocks}) tpb = \left(\frac{N - o}{tpb - o}\right) tpb
    \label{eq:operational-threads}
\end{equation}

Equation \eqref{eq:operational-threads} gives a sense of the amount of work required per cycle, which grows as overlap increases. Therefore, it is only beneficial to perform overlap as long as the number of cycles decreases. Once the number of cycles stagnates, it is no longer worth increasing amount of overlap between subdomains.

For each value of $k$ tested, Table \ref{tab:1d-optimal-results} shows the optimal overlap for minimum time and corresponding speedup of the shared memory algorithm relative to the classic GPU implementation. The results illustrate that the optimal number of overlapping points increases as the number of subiterations is increased. This is expected because increasing $k$ causes more DOFs near the subdomain edges to have poor values due to lack of up to date boundary values (so more overlap is necessary to improve these DOFs). The best performance is achieved when setting $k = 16$ with 4 interior points overlapping between subdomains. This corresponds to an eight times speedup compared to the classic GPU performance. The best performance was previously achieved using $k = 8$, but now higher $k$ values yield the best speedup because poor edge values (associated with high $k$) can be improved with overlap.

The numerical test results demonstrate that the convergence of Jacobi iteration for 1D problems can be accelerated using shared memory. Although the initial results suggested a four times speedup for this problem (resulting from tuning the number of subiterations), introducing overlap to improve edge point values results in an eightfold speedup. The optimal setting for the number of subiterations is about half the number of interior points in a subdomain. Furthermore, a small amount of overlap is enough to decrease the number of cycles dramatically. This shared memory approach can be extended to two dimensional problems for similar acceleration.

\begin{table}[htbp!]
    \centering
    \begin{tabular}{|c|c|c|c|c|c|c|}
    \hline
    Number of Subiterations & 4 & 8 & \textbf{16} & 32 & 64 & 128 \\
    \hline
    Optimal Overlap & 2 & 4 & \textbf{4} & 8 & 10 & 12 \\
    \hline 
    Optimal Speedup & 3.09 & 5.74 & \textbf{8.00} & 7.50 & 5.84 & 3.75 \\
    \hline
    \end{tabular}
    \caption{Speedup from classic GPU to shared memory approach for various subiteration values using overlapping subdomains. The optimal overlap increases with the number of subiterations performed. An eight times speedup is achieved when $k = 16$ and 4 interior points overlap between each subdomain.}
    \label{tab:1d-optimal-results}
\end{table}

\section{Approaches for Jacobi Iteration in 2D Problems}
We extend the shared memory algorithm presented in Section \ref{sec:1d-shared-algorithm} to the two-dimensional case. Consider a PDE discretized in a structured 2D domain. A model 2D PDE is one with second order derivatives, which discretized using a 2nd order central difference scheme results in a linear system where the matrix has pentadiagonal structure as shown below ($a,d,c,e,f$ correspond to different constants).

\[
A = 
\begin{pmatrix}
d & c & & f & & \\
a & d & c & & \ddots & \\
 & \ddots & \ddots & \ddots & & f \\
e &  & \ddots & \ddots & \ddots \\
 & \ddots &  & \ddots & \ddots & c \\
  &  &  e & & a & d 
\end{pmatrix}, \ \
\]

\begin{figure}[htbp!]
    \centering
    \includegraphics[width=0.25\textwidth]{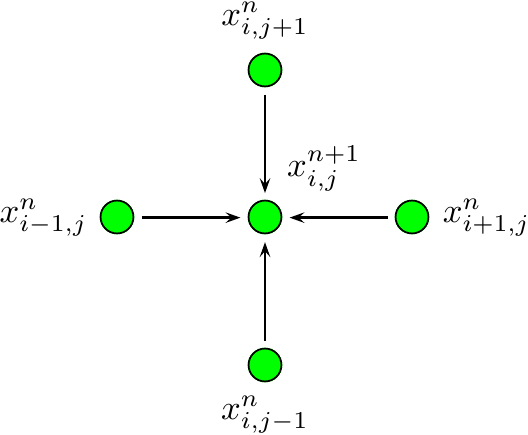}
    \caption{2D Stencil for Jacobi iteration performed on a pentadiagonal matrix. The value of a degree of freedom at the next step depends only on its neighbors' values at the current step.}
    \label{fig:2d-stencil}
\end{figure}

Applying the Jacobi iterative method to the $ij$th degree of freedom in the grid ($i$ and $j$ correspond to indices in the spatial $x$, $y$ domain) results in an update equation of the form in Equation \eqref{eq:2d-jacobi-stencil}. The updated value of degree of freedom with indices $i$, $j$ depends only on its neighbor values at the current iteration. This is depicted in Figure \ref{fig:2d-stencil}.

\begin{equation}
    x_{ij}^{(n+1)} = \frac{1}{d} \left(b_{ij} - a x_{i-1,j}^{(n)} - c x_{i+1,j}^{(n)} - e x_{i,j-1}^{(n)} - f x_{i,j+1}^{(n)} \right)
    \label{eq:2d-jacobi-stencil}
\end{equation}

\subsection{GPU Approach with Shared Memory}
\label{sec:shared-algorithm-2d}

\begin{figure}[htbp!]
    \centering
    \begin{subfigure}[b]{0.45\textwidth}
        \includegraphics[width=\textwidth]{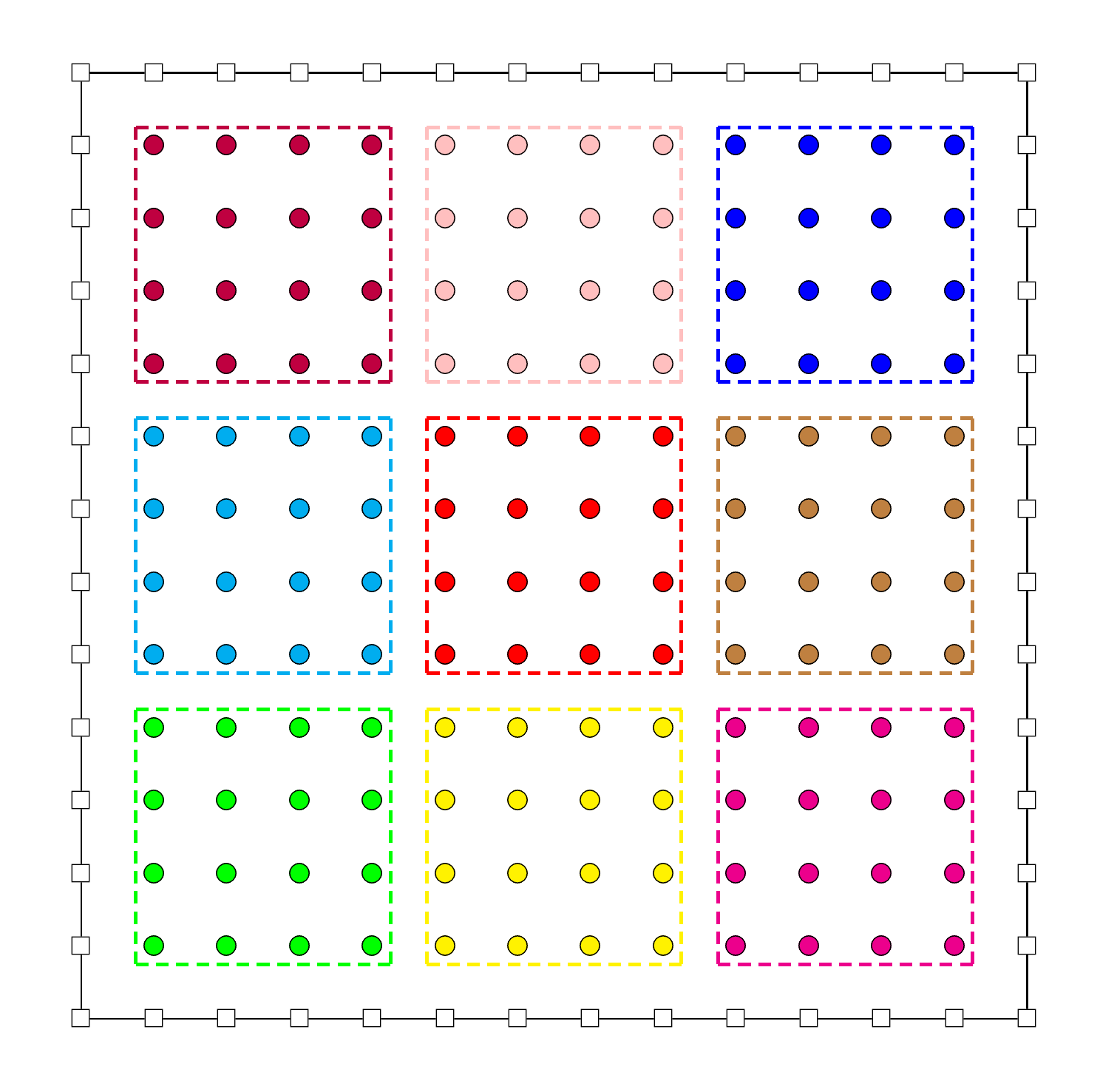}
        \caption{2D Domain and Subdomains}
        \label{fig:2d-grid-colored}
    \end{subfigure}
    \hfill
    \begin{subfigure}[b]{0.45\textwidth}
        \includegraphics[width=\textwidth]{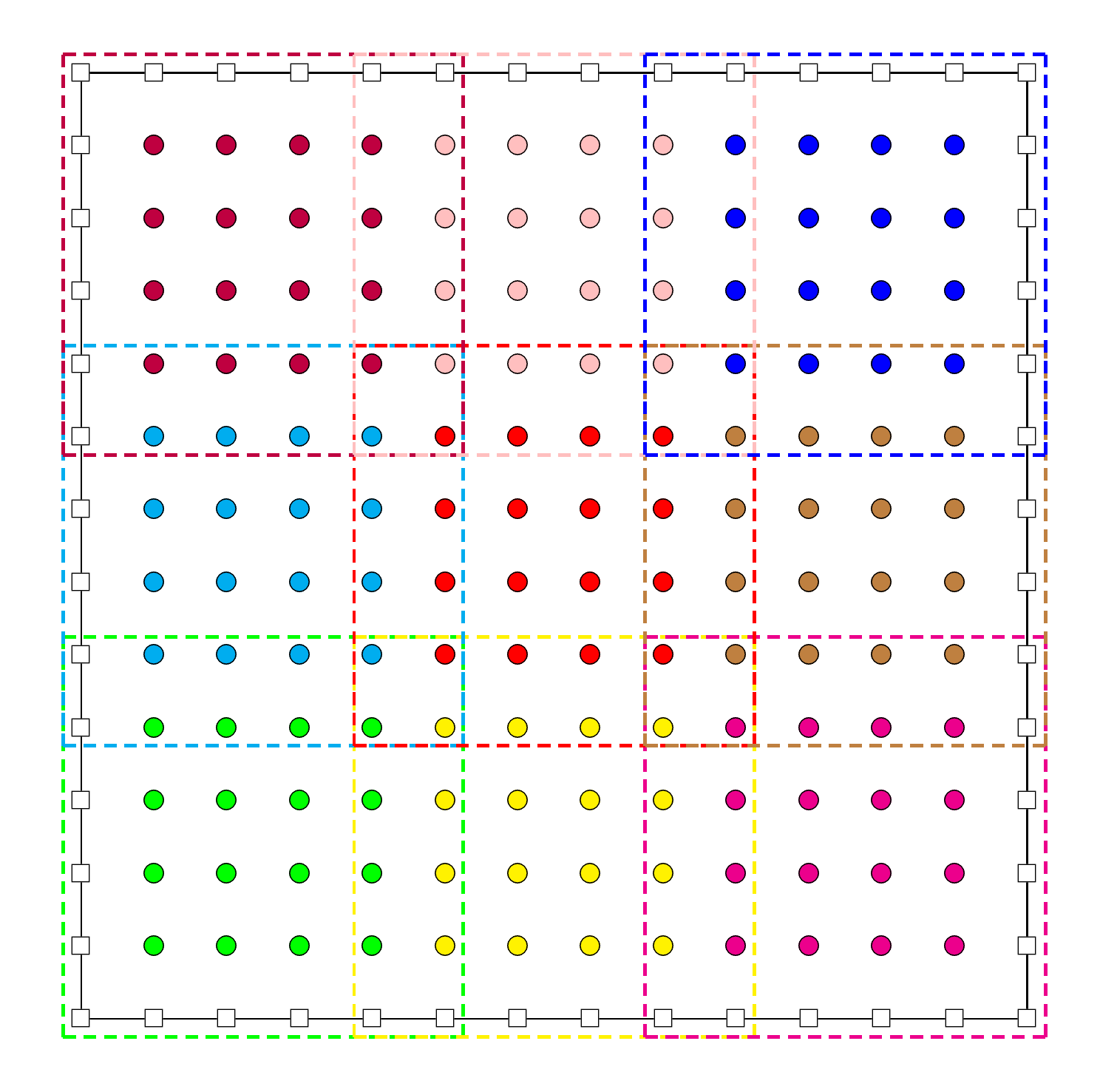}
        \caption{2D Augmented Subdomains}
        \label{fig:2d-stencil-partitioned}
    \end{subfigure}
    \caption{Illustration of 2D Domain which is partitioned into several subdomains. The augmented subdomains (shown on the right with dashed lines identifying each subdomain) are copied from global memory to the shared memory of different GPU blocks. Each GPU block will perform updates for one subdomain.}
\end{figure}

The approaches to solving a 2D PDE using Jacobi iteration on the CPU and the GPU are analogous to the 1D case. For the shared memory approach, we must utilize a domain decomposition strategy. A structured 2D domain can be subdivided into smaller subdomains which can be allocated to the shared memory of many GPU blocks and updated independently. Figure \ref{fig:2d-grid-colored} illustrates a 2D domain with 12 by 12 interior points partitioned into smaller 4 by 4 subdomains which would be updated by 9 GPU blocks (each color represents the points handled by a distinct GPU block). The square points correspond to Dirichlet boundary conditions.

To prevent threads from being idle, we ensure that every thread updates one interior point in the subdomain. To achieve this, the number of interior points in a subdomain is set to the number of threads per block in the $x$ direction (denoted by $\texttt{blockDim.x}$) by the number of threads per block in the $y$ direction (denoted by $\texttt{blockDim.y}$). Updating the points on the edges of each subdomain requires information from the edges of neighboring subdomains. To prevent the need for communication of neighbors at every step, we simply augment each subdomain by one point in all dimensions as shown in Figure \ref{fig:2d-stencil-partitioned}. Therefore, the subdomains to be copied to shared memory have size $\texttt{blockDim.x} + 2$ by $\texttt{blockDim.y} + 2$ where the inner $\texttt{blockDim.x}$ by $\texttt{blockDim.y}$ points will be updated using the Jacobi update equation.

The first step of the 2D shared memory algorithm is to copy the subdomain data from global memory to the shared memory of the GPU block which will perform the updates. The $\texttt{blockDim.x}$ by $\texttt{blockDim.y}$ threads in each block can be used to copy the $\texttt{blockDim.x}$ + 2 to $\texttt{blockDim.y}$+2 subdomains to shared memory (each thread must transfer at least one point). After the copy to shared memory, the interior points of the subdomain can be advanced using the stencil update in Equation \ref{eq:2d-jacobi-stencil}. Rather than perform just one update, we can perform multiple subiterations similar to the 1D case. Each thread is responsible for updating one DOF. After the updates are complete, the interior points of the subdomains can be copied back from shared memory to their appropriate positions in global memory. At this point, the most up to date solution lies in global memory and the shared memory can be terminated. 

This completes one cycle of the shared memory algorithm in 2D. The main parameters of this algorithm are the number of threads per block in the $x$-direction ($\texttt{blockDim.x}$) and $y$-direction ($\texttt{blockDim.y}$), which control the dimensions of the subdomains, as well as the number of subiterations performed within shared memory (denoted by $k$ as before). The algorithm should be implemented as a single GPU kernel so that shared memory is not terminated until the end. In summary, each cycle involves the following three steps (similar to the 1D case):
\begin{enumerate}
\item Partition the gridpoints between the different GPU blocks and copy values from global memory to shared memory. Each subdomain has size \\ $\texttt{blockDim.x} + 2$ by $\texttt{blockDim.y} + 2$.
\item Update the interior $\texttt{blockDim.x}$ by $\texttt{blockDim.y}$ points in each GPU block by performing $k$ Jacobi iterations. 
\item Update the global memory solution array by copying the interior $\texttt{blockDim.x}$ by $\texttt{blockDim.y}$ values from shared memory to their appropriate positions in global memory.
\end{enumerate}

We can compute the amount of shared memory required for the 2D algorithm. Since Jacobi iteration requires two storage containers, we must allocate enough data for twice the subdomain size of \\ $\texttt{(blockDim.x + 2) * (blockDim.y + 2)}$. Furthermore, the right hand side values corresponding to the interior subdomain DOFs can also be copied to shared memory to improve performance (as the right hand side appears in the update).  This requires another $\texttt{blockDim.x} * \texttt{blockDim.y}$ array. Therefore, the total amount of shared memory required is then $2 * (\texttt{blockDim.x + 2}) * (\texttt{blockDim.y + 2}) +  (\texttt{blockDim.x} * \texttt{blockDim.y}) $. This must be multiplied by the size of the representation (4 bytes for floats, 8 bytes for doubles).

One issue that can arise in shared memory applications is the problem of bank conflicts, which occur when threads in a given warp access non-contiguous data in shared memory. In this case, some fraction of the shared memory accesses are serialized which can significantly reduce memory bandwidth. To avoid performance loss, all warps should access contiguous data. In the 2D problem, each row of interior points in a subdomain lies in contiguous memory, but each subsequent row of interior points is separated from the previous row by two boundary points (due to the actual layout in memory as a 1D array). Therefore, we should ensure that each warp updates a single row. This is done by setting the threads per block value in the $x$ direction to 32, so that each row has 32 interior points (contiguous in memory) which will be updated by the 32 threads in a given warp.

\subsection{Effect of Subiterations}
We study the performance of the shared memory algorithm in solving the pentadiagonal system arising from discretization of the 2D Poisson equation in the domain $x \in [0,1] \times y \in [0,1]$, with Dirichlet boundary conditions
\begin{equation}
    - \left( \frac{\partial^2 u}{\partial x^2} + \frac{\partial^2 u}{\partial y^2} \right) = f(x,y), \ u(0,y) = u(1,y) = u(x,0) = u(x,1) = 0
    \label{eqn:2d-poisson}
\end{equation}
The pentadiagonal matrix in our linear system $Ax = b$ is given by
\[
A = 
\begin{pmatrix}
d & a & & c & & \\
a & d & a & & \ddots & \\
 & \ddots & \ddots & \ddots & & c \\
c &  & \ddots & \ddots & \ddots \\
 & \ddots &  & \ddots & \ddots & a \\
  &  &  c & & a & d 
\end{pmatrix}, \ \
\]
where
\begin{align*}
    a = -\frac{1}{(\Delta x)^2}, \ \
    c = -\frac{1}{(\Delta y)^2}, \ \
    d = \frac{2}{(\Delta x^2 + \Delta y^2)}
\end{align*}
The elemental Jacobi iterative update for the 2D Poisson problem is:
\begin{align}
    x_{i,j}^{(n+1)} &= \frac{1}{\frac{2}{\Delta x^2} + \frac{2}{\Delta y^2}} \left[b_{i,j} + \frac{1}{\Delta x^2} \left(x_{i-1,j}^{(n)} + x_{i+1,j}^{(n)} \right) + \frac{1}{\Delta y^2} \left(x_{i,j-1}^{(n)} + x_{i,j+1}^{(n)} \right) \right] \\
    &= \frac{b_{i,j} (\Delta x)^2 + x_{i-1,j}^{(n)} + x_{i+1,j}^{(n)} + x_{i,j-1}^{(n)} + x_{i,j+1}^{(n)}}{4} \ \ \ \text{if} \ \ \Delta x = \Delta y
\end{align}

We consider a domain containing $N_x = 1024$ by $N_y = 1024$ interior grid points in the $x$ and $y$ directions and measure the time to reduce the $L_2$ residual norm of the initial solution by a factor of 1e-4. This problem is analogous to the previous 1D problem in terms of computational complexity, and requires full utilization of the GPU. The entries of the initial solution and the right hand side vector are set to one. A TITAN V GPU is used for the numerical tests, which are performed using double precision.

The number of threads per block in the $x$ direction is set to 32 to avoid shared memory bank conflicts. For the classic GPU approach, the threads per block value in the $y$ direction is varied between $4, 8, 16, 32$ and the best time is recorded. For the shared memory approach, the threads per block value in the $y$ direction is set to 32 so that the problem domain is split into subdomains with 32 by 32 interior points. The amount of shared memory required using these specifications is 26.688 kB per block (less than the maximum allowable value of 48 kB, although doubling either dimension of the subdomain would cause us to exceed this value). The number of subiterations is set to the following values (the same values as in the 1D problem): $k = 4, 8, 16, 32, 64, 128$.

\begin{figure}[htbp!]
    \centering
    \includegraphics[width=0.6\textwidth]{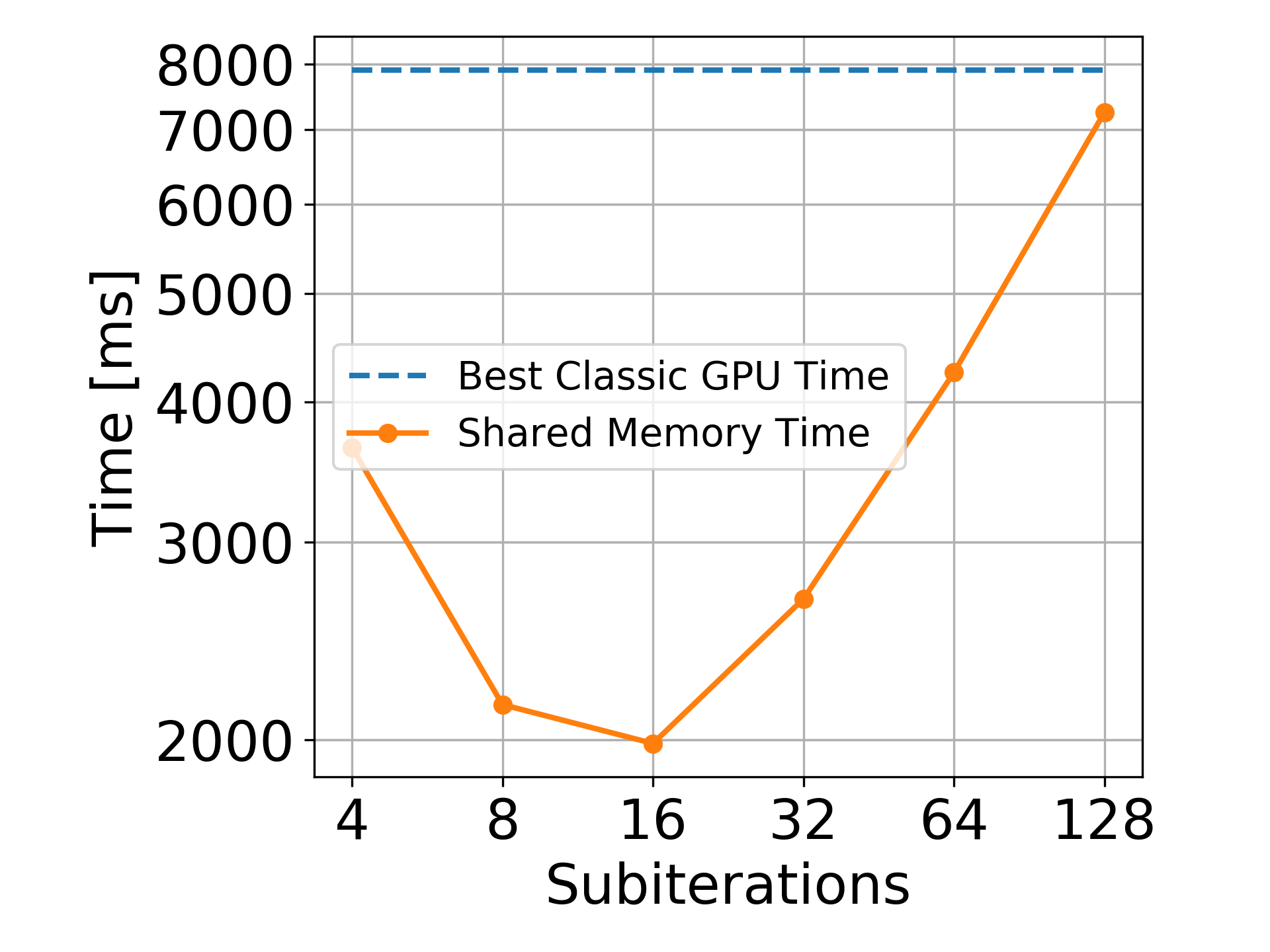}
    \caption{Comparison of time required for classic GPU and shared memory approaches to reduce the residual of the solution by 1e-4 for a 2D problem size of $N_x = 1024$ by $N_y = 1024$. The shared memory approach outperforms the best classic GPU time in all cases, and the best performance is achieved when $k = 16$.}
    \label{fig:gpu-shared-2D}
\end{figure}

Figure \ref{fig:gpu-shared-2D} shows a comparison of the time required for the classic GPU and shared memory approaches to decrease the initial residual by a factor of 1e-4. The time required for all data transfers between the host and device are also included in the results. The best classic GPU time measured was $7901.1$ ms for the various thread per block specifications we experimented with. The shared memory approach outperforms the best classic GPU performance for all cases. The best performance is achieved when $k = 16$. The time grows linearly as the number of subiterations is increased past this point.

Table \ref{tab:gpu-vs-shared-speedup-2d} shows the speedup obtained when comparing the best classic GPU performance to the shared performance for each of the subiteration values explored. A nearly four times speedup is achieved using the shared memory implementation (similar to the initial 1D speedups) when $k = 16$, demonstrating that shared memory can improve the convergence of Jacobi iteration for structured 2D problems. 

\begin{table}[htbp!]
    \centering
    \begin{tabular}{|c|c|c|c|c|c|c|}
    \hline
    Number of Subiterations & 4 & 8 & \textbf{16} & 32 & 64 & 128 \\
    \hline
    Best GPU to Shared Speedup & 
    2.17 & 3.67 & \textbf{3.98} & 2.96 & 1.86 & 1.09 \\
    \hline
    \end{tabular}
    \caption{Speedup from classic GPU to shared memory approach for various subiteration values used in the 2D problem. A speedup of nearly four is achieved when setting $k = 16$.}
    \label{tab:gpu-vs-shared-speedup-2d}
\end{table}

\subsection{Effect of Overlapping Subdomains in 2D}
The shared memory approach improves the performance of classic Jacobi iteration on the GPU by allowing many updates to be performed without requiring global memory accesses every step. However, this comes at the cost of lacking up to date boundary values every subiteration. In the 2D problem, the left, right, top and bottom boundary edges are only up to date at the beginning of a cycle but not for all $k$ subiterations. This causes DOFs close to these subdomain edges to rely on old boundary data and have poor solution values which contribute greatly to the residual norm (especially for larger values of $k$). Improving the interior values near subdomain edges (especially interior points at the corners) can greatly improve the convergence of our algorithm. This can be done by extending the overlapping approach in Section \ref{sec:overlap} to 2D problems. In the 2D case, sets of points can overlap between subdomains in both the $x$ and $y$ directions. Therefore, a DOF can be allocated to and updated within multiple subdomains. However, the subdomain in which the local position of the DOF is further from a subdomain edge in both the $x$ and $y$ directions should contribute the final value at the end of a cycle. This is a natural extension of the 1D overlapping approach to the 2D case.

\begin{figure}[htbp!]
    \centering
    \begin{subfigure}[b]{0.6\textwidth}
        \includegraphics[width=\textwidth]{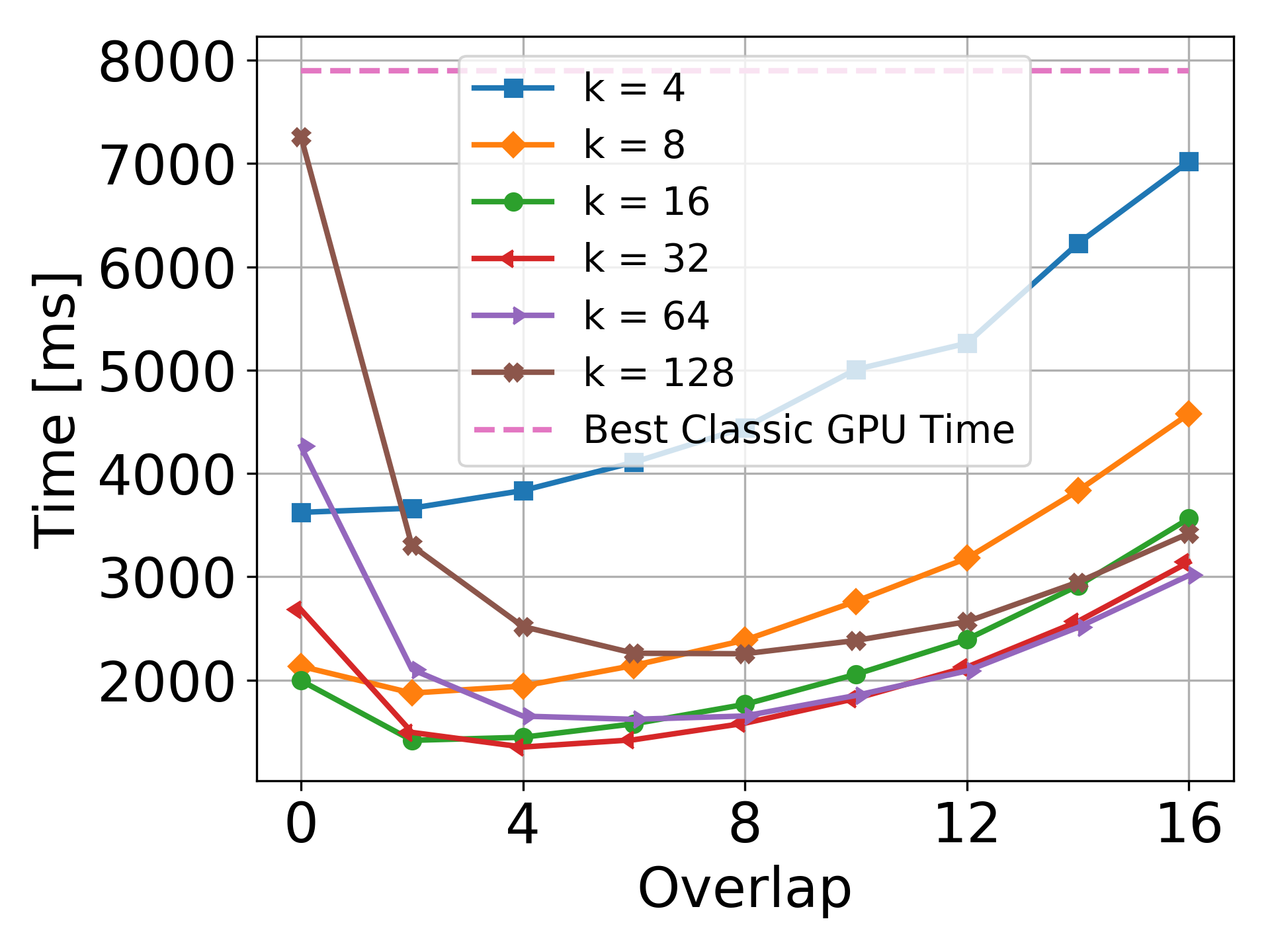}
        \caption{Time}
        \label{fig:subiteration-time-2d}
    \end{subfigure}
    \hfill
    \begin{subfigure}[b]{0.6\textwidth}
        \includegraphics[width=\textwidth]{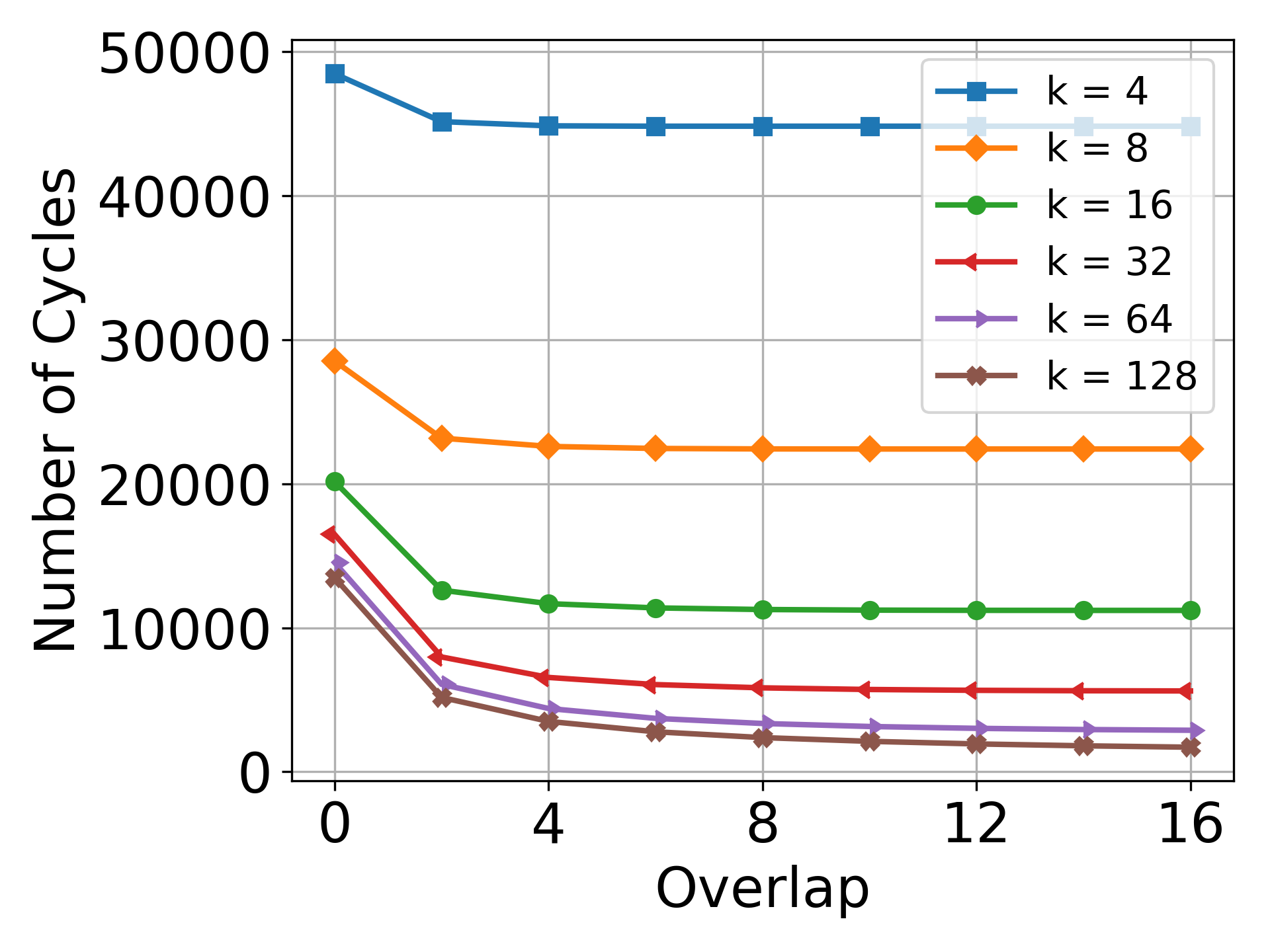}
        \caption{Number of Cycles}
        \label{fig:subiteration-cycles-2d}
    \end{subfigure}
    \caption{Time and Number of Cycles required for convergence as a function of overlap for different number of subiterations $k$ for the 2D problem. The behavior is similar to the 1D case. The best performance is achieved with $k = 16$.}
\end{figure}

We explore the performance of the 2D shared memory algorithm with overlap. As before, the thread per block values are set to 32 in both directions. In this case, a subdomain can have up to 30 interior points overlap with its neighboring subdomains in both directions. While it is permissible to have a different amount of overlap in the $x$ and $y$ directions, the symmetry of the domain and subdomain sizes suggests that the optimal overlap amount would be the same in both directions. 
Figures \ref{fig:subiteration-time-2d} and \ref{fig:subiteration-cycles-2d} show the time and number of cycles required for the shared memory approach to converge as the overlap in both directions is increased. The same subiteration values are used as before. Figure \ref{fig:subiteration-cycles-2d} shows that an initial amount of overlap can significantly drop the number of cycles required for convergence, especially when the number of subiterations is large. The corresponding time also decreases, except in the case when $k = 4$, as the decrease in the number of cycles is not enough to outweigh the extra work incurred due to overlap. As the number of overlapping points increases further, the number of cycles saturates. However, the time continues to increase because larger overlapping values require more GPU resources and computational time per cycle.

We conduct a similar analysis to the 1D case to determine the amount of work required per cycle as a function of the overlap. The 1D analysis suggested that the number of operational blocks was given by:
\begin{equation}
    \text{Operational blocks} = \frac{N - o}{tpb - o}
\end{equation}
For the 2D case, assume a domain with interior dimensions $N_x$ by $N_y$, with subdomains with interior dimensions of $tpb_x$ by $tpb_y$. If the overlap in the $x$ and $y$ direction are given by $o_x$ and $o_y$, the total number of operational blocks required is then
\begin{equation}
    \text{Operational blocks} = \left( \frac{N_x - o_x}{tpb_x - o_x} \right) \left( \frac{N_y - o_y}{tpb_y - o_y} \right)
\end{equation}
The number of operational threads is then the number of operational blocks times the threads per block in the $x$ and $y$ directions
\begin{equation}
    \label{eq:op-threads}
    \text{Operational threads} = \left( \frac{N_x - o_x}{tpb_x - o_x} \right) \left( \frac{N_y - o_y}{tpb_y - o_y} \right) tpb_x tpb_y
\end{equation}
Assuming that the parameters in the $x$ and $y$ directions are equivalent ($N_x = N_y = N$, $tpb_x = tpb_y = tpb$, $o_x = o_y = o$ as is true in our problem) Equation \eqref{eq:op-threads} simplifies to
\begin{equation}
    \label{eq:operational-threads-2d}
    \text{Operational threads} = \left( \frac{N - o}{tpb - o} \right)^2 tpb^2
\end{equation}

Equation \ref{eq:operational-threads-2d} shows the amount of work or computational resources required per cycle based on the domain size, threads per block and overlap. Increasing the overlap increases the amount of work required per cycle, so overlap is useful if it decreases the number of cycles needed for convergence enough to outweigh the extra work it requires. Once the number of cycles saturates with overlap, it is no longer useful to perform further overlap.

Table \ref{tab:gpu-vs-shared-speedup-2d-optimal} shows the optimal overlap value and corresponding speedup of the shared memory algorithm relative to the classic GPU implementation for the 2D problem, for each of the subiteration values used. As before, the optimal overlap increases with the number of subiterations. A nearly six times speedup is achieved using $k = 32$ with 4 overlapping points in the $x$ and $y$ directions.
\begin{table}[htbp!]
    \centering
    \begin{tabular}{|c|c|c|c|c|c|c|}
    \hline
    Number of Subiterations & 4 & 8 & 16 & \textbf{32} & 64 & 128 \\
    \hline
    Optimal Overlap in $x$ and $y$ & 0 & 2 & 2 & \textbf{4} & 6 & 8 \\
    \hline
    Optimal Speedup & 2.18 & 4.22 & 5.58 & \textbf{5.84} & 4.88 & 3.50 \\
    \hline
    \end{tabular}
    \caption{Speedup from classic GPU to shared memory approach for various subiteration values used in the 2D problem, with overlapping subdomains. The best speedup (nearly sixfold) is achieved when $k = 32$ with 4 overlapping points between subdomains in both directions.}
    \label{tab:gpu-vs-shared-speedup-2d-optimal}
\end{table}
The numerical test results demonstrate that the convergence of Jacobi iteration for 2D problems can also be accelerated using shared memory. Tuning the number of subiterations initially resulted in a four times speedup over the classical approach. However, introducing overlap to improve edge values augmented the speedup further, resulting in an almost six times speedup. The optimal number of subiterations was equal to the number of interior points in a subdomain along a single dimension. Furthermore, introducing a small amount of overlap in both directions was enough to decrease the number cycles dramatically. 

\section{Conclusion}
In this work, we have developed a hierarchical Jacobi solver for 1D and 2D structured problems that utilizes shared memory to improve performance. The approach involves partitioning the domain into multiple subdomains which are handled by GPU blocks. Relative to a classical GPU implementation of the Jacobi iterative method, the shared memory approach resulted in an eight times speedup in convergence for the 1D case and a nearly six times speedup for the 2D case. This speedup was achieved by exploring optimal parameters for the number of subiterations performed in shared memory and investigating the effect of overlapping subdomains (which accelerated convergence by improving DOF values near subdomain edges). For both the 1D and 2D test problems, setting the number of  subiterations on the order of the subdomain dimension along with a small amount of overlap resulted in the optimal performance of the shared memory algorithm. This hierarchical Jacobi solver is very suitable for parallel architectures, and has the potential to accelerate multigrid solvers which utilize this approach as a smoother on high performance clusters. Overall, this work demonstrates the need to adopt classical algorithms to emerging hardware in a way that achieves both high computational throughput as well as efficient memory access. 

\appendix
\clearpage
\section{Sample Pseudocode for 1D Shared Memory Algorithm} 
We provide a sample CUDA C++ style code to illustrate the components of the shared memory algorithm in 1D from an implementation standpoint. The code resembles the implementation of our GPU kernel which corresponds to one cycle of the shared memory algorithm. In this first step, the subdomain DOF data is copied from global memory to 2 containers in shared memory (corresponding to containers for the current and updated values in Jacobi iteration). The right hand side values for DOFs corresponding to our subdomain are also copied to a third section of shared memory. In the second step, Jacobi subiterations are performed in shared memory. We have two storage containers for Jacobi iteration, with x0 referring to the current solution and x1 referring to the updated solution in every step. In step 3, the updated values of all interior points in shared memory are copied back to their respective global memory positions.

\lstset{language=C++}  
\begin{lstlisting}

/* Define shared memory */
extern __sharedMemory__ double sharedMemory[];

/* Define several constants */
int subdomainSize = blockDim.x + 2;
int i = threadIdx.x;
int I = threadIdx.x + blockDim.x * blockIdx.x;

/* STEP 1 - Copy data from global memory to shared memory */
sharedMemory[i] = globalMemory[I]; // 1st container
sharedMemory[i + subdomainSize] = globalMemory[I]; // 2nd contain.
sharedMemory[i + 2 * subdomainSize] = rhs[I + 1]
if (threadIdx.x == 0 || threadIdx.x == 1) {
    sharedMemory[i + blockDim.x] = globalMemory[I + blockDim.x];
    sharedMemory[i + blockDim.x + subdomainSize] = ...
    globalMemory[i + blockDim.x]
}

/* STEP 2 - Perform subiterations within shared memory */
i = threadIdx.x + 1
double * x0 = sharedMemory; // 1st container
double * x1 = sharedMemory + subdomainSize; // 2nd container
double * sharedrhs = sharedMemory + 2 * subdomainSize
for (int k = 0; k < numSubiterations; k++) {
    int left_value = x0[i-1];
    int right_value = x0[i+1];
    int rhs_value = sharedrhs[i]
    x1[i] = jacobiUpdate(left_value, right_value, rhs_value)
    __syncthreads();
    swap(x0, x1);
}

/* STEP 3 - Transfer from shared memory to global memory */
globalMemory[I+1] = sharedMemory[i]

\end{lstlisting}

\clearpage
\printbibliography

\end{document}